\documentclass{jetpl}
\usepackage{epsfig}

\twocolumn

\def \sr{Schr\"{o}dinger~}
\def\fg#1 {Fig.\,\ref{#1} }
\def\bra#1{{\langle #1|}}
\def\ket#1{{| #1\rangle}}
\def\bracket#1#2 {\mathinner{\langle{#1}|{#2}\rangle}}

\begin{document}

\title{ Thermodynamic magnetization of two-dimensional electron gas measured over wide range of densities}
\rtitle{ Thermodynamic magnetization of two-dimensional electron
gas measured over wide range of densities}

\sodtitle{ Thermodynamic magnetization of two-dimensional electron
gas measured over wide range of densities}

\author{M.~Reznikov$^{a}$\/\thanks{e-mail: reznikov@tx.technion.ac.il}, A.~Yu.~Kuntsevich$^{b}$, N.~Teneh$^{a}$, V.~M.~Pudalov$^b$}

\rauthor{M.~Reznikov, A.~Yu.~Kuntsevich, N.~Teneh, V.~M.~Pudalov}

\sodauthor{Reznikov, Kuntsevich, Teneh, Pudalov}

\address{$^{a}$
Solid State Institute, Technion, Haifa, 3200 Israel\\
$^{b}$ P.N. Lebedev Physical Institute, 119991 Moscow, Russia}

\abstract{ We report measurements of $\partial m/\partial n$ in Si
MOSFET, where $m$ is the magnetization of the two-dimensional
electron gas and $n$ is its density.  We extended the density
range of measurements from well in the metallic to deep in  the
insulating region. The paper discusses in detail the conditions
under which this extension is justified, as well as the
corrections one should make to extract $\partial m/\partial n$
properly. At low temperatures, $\partial m/\partial n$ was found
to be strongly nonlinear already in weak magnetic fields, on a
scale much smaller than the characteristic scales, expected for
interacting two-dimensional electron gas. Surprisingly, this
nonlinear behavior exists both in the dielectric, and in the
metallic region. These observations, we believe, provide evidence
for strong coupling of the itinerant and localized electrons in
Si-MOSFET.}

\date{\today}

\maketitle

Magnetic properties of strongly-interacting electron gas have long
been a subject of intensive theoretical and experimental
investigations. Coulomb interaction through the exchange term
favors parallel spins, and therefore leads to ferromagnetism,
whenever is strong enough. The strength of the interaction is
determined by the ratio between the typical Coulomb and kinetic
energies, and  customarily characterized by the ratio between the
typical inter-electron distance $\rho$,  $\rho=1/\sqrt{\pi n}$ in
two dimensions, and the effective Bohr radius: $r_s=\rho/a_B^*$
\cite{AFS}. When the carrier density $n$ decreases, the relative
effect of the interactions becomes stronger, contrary to the naive
expectations. At sufficiently low density a clean system of
itinerant electrons was predicted to become ferromagnetic, the
phenomenon called Stoner instability \cite{Stoner}; though in two
dimensions it can occur only at zero temperature.

A real system of itinerant electrons is always disordered to some
extent. Interactions, in the limiting case of strong disorder,
favor antiparallel arrangements of the neighboring spins. The
ground state of a pair of localized spins is a singlet, much like
a Hydrogen molecule. A system of many localized spins preserves
the tendency to antiferromagnetic order, see~\cite{BhatReview} for
review. Intensive investigation of the magnetic properties of the
doped semiconductors, particularly phosphorus doped Si  in
80s~\cite{BhatReview}, lead to substantial understanding of the
interplay between interactions and disorder. The observed
divergency of the susceptibility with decreasing temperature  at
the low-density side of the metal-insulator transition was well
understood; what remained unanswered, is the divergent
susceptibility at the high-density side of the transition.

Interactions in 2D systems are more important than in 3D, which
explains the wealth of experimental observations and difficulty in
their explanation.  Unfortunately, extremely small number of
electrons hinders thermodynamic measurements of 2D systems.  On
the other hand, two-dimensional gated structures, particularly
Si-MOSFETs, provide unique possibility of gradually changing the
electron density, and, thus, interaction strength. Interest in the
magnetic properties of 2DEG was sparked by the observation of
strong suppression of conductivity in Si-MOSFETs by in-plane
magnetic field~\cite{simonyan}.  The strong magnetoresistance was
interpreted in~\cite{SKDK,VZ2} as a quantum phase transition into
a ferromagnetic state at the density of the metal-insulator
transition, $n_c$. This interpretation was contested
in\,\cite{pudalovsdh,PGK2001} on the basis of Shubnikov-de\,Haas
measurements, and  in\,\cite{Prus2003} on the basis of
thermodynamic magnetization measurements.

In the thermodynamic method, developed in\,\cite{Prus2003},
recharging current, generated between the 2DEG and the gate in
response to modulation of external magnetic field $B$, is used to
determine the derivative of the 2DEG chemical potential $\mu_{2D}$
with respect to the field. This derivative, by virtue of the
Maxwell relation $\partial \mu/\partial B=-\partial m/\partial n$,
can be translated into the derivative of the magnetic moment $m$
at a given field with respect to the density $n$. One can, in
principle, integrate $-\partial m/\partial n$ over $n$ to get $m
(n)$, assuming  $m$ is known at a certain density. In
Refs.\,\cite{Prus2003,shashkintd} the integration constant was
taken from transport measurements at some high density value.

It would be desirable to abandon this assumption by
straightforward integration from $n=0$, since at zero density the
magnetic moment of an electron gas is zero by definition. Such
approach, however, requires the measurements to be performed down
to very low carrier densities, lower than were accessible earlier.
In the current paper we extended the thermodynamic measurements
down to densities as low as $0.3\,n_c$, deep in the insulating
region. This extension enabled us to reveal strong unexpected
nonlinearity of $\partial m/\partial n$ in weak magnetic fields,
which, surprisingly, exists also in the metal phase.

\begin{figure}
\centerline{\psfig{figure=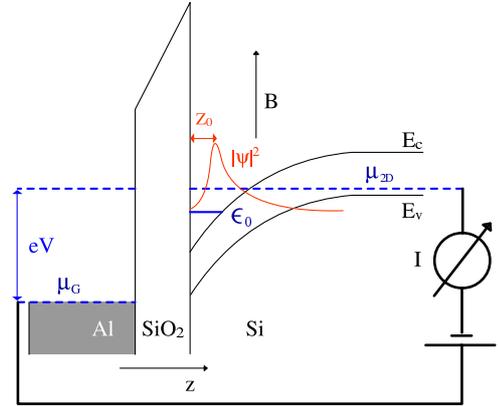,width=180pt}}
\begin{minipage}{3.2in}
\caption{Fig.\,1. A sketch of the experimental setup, superimposed
on the band diagram of a Si MOSFET sample; $z$-direction is
perpendicular to the 2DEG plane}\label{Fig:Illustration}
\end{minipage}
\end{figure}

Before presenting the experimental results, we first discuss in
detail the way we extended the measurements, as well as the
corrections which should be applied to properly interpret the
data. Consider a system consisting of a 2DEG connected through an
ohmic contact and a battery, which provides the gate voltage $V$,
to the gate, as shown in Fig.\,\ref{Fig:Illustration}. The free
energy of the system can be written as follows:
\begin{equation}
\label{F}
f=f_G+f_{2D}-enV +\frac{e^2n^2}{2c_0},
\end{equation}
where $f_G$ and $f_{2D}$ are the free energies of the gate and the 2DEG, respectively, and the last term describes the electrostatic interaction between the 2DEG and the gate electrons.
The capacitance $c_0$, defined by Eq.\,(\ref{F}), differs from the geometric one by a factor of  $\approx (1+\bar z_0/ d_{\rm ox})^{-1}$; here $d_{\rm ox}\approx 190$\,nm is the oxide thickness  for the studied Si-MOSFETs, and $\bar z_0 \approx 3.5$\,nm is the average distance of the 2D electron layer from the Si/SiO$_2$ interface.

The minimum of $f$ determines the equilibrium density of the 2DEG. Differentiating Eq.\,(\ref{F}) with respect to $n$ under the constrain of zero net charge of the system (the variation of the  electron density at the gate $\delta n_g=-\delta n$) gives the familiar expression:
\begin{equation}\label{Eq:n} n=\frac{c_0}{e}\left(V-\frac{\mu_{2D}-\mu_G}{e}\right)
\end{equation}
where $\mu_{2D}$ and $\mu_G$ are the chemical potentials of
the electrons in the 2DEG and the gate respectively.  We
ignored $c_0$ dependence on the density; accounting for it leads
to small corrections for both $n$, and $\tilde c$ defined below,
of the order of $({\partial c_0}/{\partial n})({n}/{c_0})$.

A change of an external parameter shifts the equilibrium density
$n$. We assume $\mu_G$, the chemical potential of a  thin Al
film, to be independent of $B$ and $n$. Consider first the
density response  on a variation of $V$, i.e. $dn/dV\equiv \tilde
c /e$, which defines the gate-to-2DEG capacitance $\tilde
c$\cite{commentonC}:
\begin{equation} \label{Eq:dn} \tilde
c=c_0\left(1+\frac{c_0}{e^2}\frac{\partial \mu_{2D}}{\partial
n}\right)^{-1}
\end{equation}
As seen from Eq.\,(\ref{Eq:dn}), $\tilde c$ is slightly renormalized,  by a finite 2DEG compressibility.

Similarly, the density response to magnetic field variation is
given by:
\begin{equation}
\label{Eq:dB}
\frac{e^2}{\tilde c}\frac{d n}{d B}= -\frac{\partial \mu_{2D}}{\partial B} + \frac{e^2n}{c_0^2}\frac{\partial c_0}{\partial B}\equiv-\frac{\partial \tilde\mu}{\partial B}.
\end{equation}
By measuring the recharging current in response to the gate
voltage modulation one can get the capacitance $\tilde c$, and
then use it to extract $\partial \tilde\mu/\partial B$. In
Eq.\,\ref{Eq:dB} we defined $\tilde\mu$, which contains an
additional contribution proportional to ${\partial c_0}/{\partial
B}$, neglected in\,\cite{Prus2003,shashkintd}. As we shall see
below, it is indeed small for a 2DEG in Si-MOSFETs.

Since the thickness of 2DEG is finite, albeit small, an in-plane
magnetic field couples also to the orbital motion, leading to two
effects: (i) a diamagnetic shift of the spatial quantization
levels\,\cite{Prus2003,Rcndmt,comment_R_S}; (ii) a change in the
average distance $\bar z_0 $ of the electrons from the ${\rm
Si-Si0_2}$ interface due to the asymmetry of the confining
potential, see Fig.\,\ref{Fig:Illustration}. The former effect
causes a diamagnetic contribution to the chemical potential shift,
$\delta\mu_{d}$, which should be subtracted, if one is interested
in the spin contribution only: $\delta\mu_{s}=\delta
\mu_{2D}-\delta\mu_{d}$. The later effect leads to a change of the
capacitance $c_0(B)$, and results in the contribution
$\delta\mu_c$, which should be subtracted from $\delta\tilde\mu$
in order to get $\delta \mu_{2D}$: $\delta
\mu_{2D}=\delta\tilde\mu -\delta\mu_c$ with
\begin{equation}
\label{Eq:muc}
\frac{\partial\mu_c}{\partial B}=-\frac{{ne^2}}{c_0^2}\frac{\partial c_0}{\partial B}.
\end{equation}

For a non-interacting 2D gas these finite thickness (FT) contributions can be found by solving numerically the \sr equation in a triangular potential. We are interested in the energy of the lowest level $\ket 0$ of the spatial quantization, since it is the only one populated at the experimentally relevant densities. We choose Landau gauge $\vec A=(0, B(z-z_0),0)$, where $z_0=\bra 0 z \ket 0$. Such a gauge preserves the minimum of the dispersion relation $\epsilon_0(k_x,k_y)$ at $k_y=0$. Care should be taken to average the result over the in-plane $k$-vectors of the populated states to get $\bar \epsilon_0=\langle \epsilon_0\rangle_k$ and $\bar z_0=\langle z_0\rangle_k$, since components with different $k_y$ are affected differently by the field. For density-dependent confining electric field $E$ we adopted the expression\,\cite{AFS}
\begin{equation}
E=\frac{4 e\pi}{\varepsilon_{Si}}\left(n_D+\frac{11}{32}n\right),
\end{equation}
where $e\cdot n_D$ is the depletion region charge, $n_D\approx 10^{11}{\,\rm cm^{-2}}$ for our samples.
At low, compared to the Fermi energy, temperatures the diamagnetic
contribution ${\partial\mu_d}/{\partial B}=\partial^2
(n\bar\epsilon_0)/\partial n\partial B$. The capacitance
contribution, Eq.\,\ref{Eq:muc}, can be expressed through $\bar
z_0$ as:
\begin{equation}
\frac{\partial\mu_c}{\partial B}={ne^2}\frac{\partial(1/c_0)}{\partial B}=\frac{4\pi n e^2}{\varepsilon_{Si}}\frac{\partial\bar z_0}{\partial B}
\end{equation}

\begin{figure}
\begin{center}
\centerline{\psfig{figure=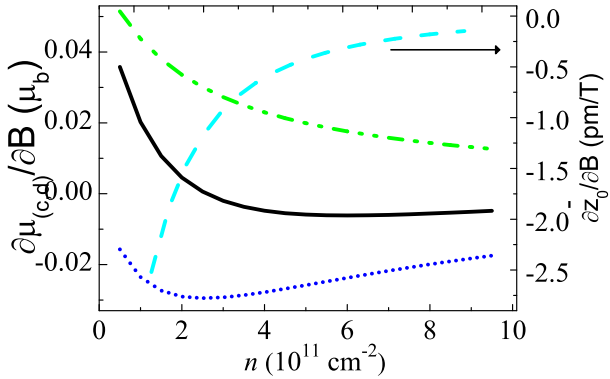, width=230pt}}
\caption{Fig.\,2. Finite thickness contributions to
${\partial\tilde \mu}/{\partial B}$ at $B=1$\,T; dashed-dotted
line  is the diamagnetic contribution ${\partial\mu_d}/{\partial
B}$, dotted line -- $\partial \mu_c/{\partial B}$, solid line --
the sum of the two. Dashed line is $\partial \bar z_0/\partial B$
(right scale, note that $\bar z_0\approx 3.5 {\rm\, nm}$).}
\label{Fig:dia}
\end{center}
\vskip-.3in
\end{figure}

\begin{figure}[h]
\begin{center}
\centerline{\psfig{figure=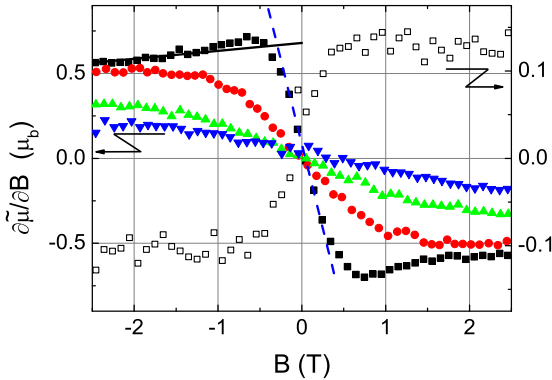, width=230pt}}
\caption{Fig.\,3. ${\partial \tilde\mu}/{\partial B}$ at
$n=5\cdot10^{10}{\rm \,cm^{-2}}$, deep in the insulating regime,
full symbols, left scale; the temperatures are: $\blacksquare$ -
1.7K $\bullet$ - 3.4K $\blacktriangle$ - 4.6K $\blacktriangledown$
- 6.8K. $\square$ - ${\partial \tilde\mu}/{\partial B}$ at
$n=2\cdot10^{11}{\rm \,cm^{-2}}$ and 1.7K, deep in the metallic
region, right scale. The slope of the dashed line gives the
derivative of susceptibility $-\partial\chi/\partial n$, and the
slope of the solid line -- estimated FT contributions, see
Fig.\,\ref{Fig:dia}.} \label{Fig:dmdB}
\end{center}
\vskip-.3in
\end{figure}

Both contributions, being the first derivatives of even functions,
are linear in $B$ at weak magnetic fields; they are shown in
Fig.\,\ref{Fig:dia}.  For $B>0$ the diamagnetic contribution is
positive, and the capacitance one is negative (it is shown in
Fig.\,\ref{Fig:dia} with negative sign). The sum of the FT
contributions is positive, therefore disregarding them would lead
to {\it underestimation} of $\partial m_s/\partial
n=-\partial\mu_s/\partial B$. For our system the resultant
contribution does not exceed $0.05 \mu_B$ at $B=1\,{\rm T}$ at low
densities and drops with $n$; therefore it is small compared to
the measured $\partial\tilde\mu/\partial B$, at least for  low
temperatures, see Fig.\,\ref{Fig:dmdB}. The FT contributions
should be more important for systems with lower effective mass in
$z$-direction, such as GaAs based heterostructures, or in higher
magnetic fields\,\cite{Prus2003}.

When the gate voltage or the magnetic field is modulated at a
sufficiently small frequency $\omega$, so that the system stays in
the thermodynamic equilibrium, the modulation leads to a
recharging current with the amplitude \begin{equation} \label{I}
(a)~~I_V=i\omega C(\omega)\delta V~~~ (b)~~I_B=-\frac{i\omega
C(\omega)}{e}\frac{\partial\tilde \mu}{\partial B}\delta B,
\end{equation}
respectively, where the capacitance $C=S\tilde c$ is proportional to the sample area $S$; Eqs.\,\ref{I} were used in~\cite{Prus2003,shashkintd}, in order to  determine the capacitance (\ref{I}a), and then to  extract ${\partial \tilde \mu}/{\partial B}$ (\ref{I}b).

However, for low electron densities, recharging of the sample is
hindered by the large contact and 2DEG resistances. When the
resistance approaches $1/\omega C$, the capacitance becomes
complex, and its magnitude drops. We show below that, under
certain assumptions, Eqs.\,(\ref{I}) can still be used for
extracting ${\partial \tilde\mu}/{\partial B}$ using the complex
capacitance, provided both $I_V$ and $I_B$ are measured at the
same conditions: frequency, magnetic field and temperature.

Recharging of the capacitor, either by modulation of the gate voltage or of the magnetic field, leads to gradients of the potential and the carrier density over the 2DEG area, and the electrochemical potential $\mu_{2D}$ becomes a function of coordinates. In addition, recharging current leads to a potential drop across the contact. The current density $\vec j$ is governed by the continuity equation:
\begin{equation}
\label{dndt}
e\frac{dn}{dt}=-\nabla\vec j=\nabla\left(\sigma\nabla\frac{\mu_{2D}}{e}\right),
\end{equation}
where $\sigma$ is, in general, a coordinate-dependent
conductivity of the 2DEG and the contact region.

Following\,\cite{Dolgopolov_skin} we assume that the system stays
in quasi-thermodynamic equilibrium, and that the characteristic
spatial scale of the electrochemical potential variations is large
compared to the oxide thickness. Then, Eq.\,(\ref{Eq:n}) with
coordinate-dependent $\mu_{2D}(r)$ is satisfied locally, the
assumption called the ``local capacitance approximation''. Under
this assumption, according to Eqs.\,(\ref{Eq:dn}) and
(\ref{Eq:dB}), both voltage and magnetic field modulation generate
variation of the density, with the only difference being $e\delta
V$ replaced with $-{\partial \tilde\mu}/{\partial B}\delta B$.
These variations enter as the source term in Eq.\,(\ref{dndt}),
whose Green's function, subject to the proper boundary conditions,
determines the current through the contact. This means that
Eqs.\,(\ref{I}) hold under local capacitance approximation, with
the same $C$ in Eqs.\,(\ref{I}a) and (\ref{I}b).

We present results obtained with the most extensively measured
Si-MOSFET sample with the peak mobility of 3.4${\rm m^2/Vs}$ at
1.7K. This sample is similar to that used in
Ref.\,\cite{simonyan}; the metal-insulator transition in it occurs
at a critical density $n_c\approx 8.5\cdot 10^{10}{\rm\,cm^{-2}}$.
The results obtained with many other samples of similar quality
were similar. Most of the measurements were done with in-plane
magnetic field modulated at frequency 6.2\,Hz  with 40\,mT
amplitude. The results scaled linearly with modulation amplitude.
Figure~\ref{Fig:dmdB} shows ${\partial \tilde\mu}/{\partial B}$ as
a function of the magnetic field. It is antisymmetric with $B$ at
6.2\,Hz, as expected for a derivative of an even function. At
higher frequencies an asymmetry develops, we believe, due to
mechanical vibrations; it was verified that the antisymmetrized
with respect to magnetic field ${\partial \tilde\mu}/{\partial B}$
is frequency-independent up to 30\,Hz. The upper measurement
frequency is limited by the EMF in wire loops, as well as by eddy
current heating.

From now on let us discuss the results for $B\geq 0$. As seen in
Fig.\,\ref{Fig:dmdB}, at the lowest temperature of 1.7K,
${\partial \tilde\mu}/{\partial B}$ is strongly  nonlinear. For
densities deep in the insulating regime, e.g. at $n=5\cdot
10^{10}{\,\rm cm^{-2}}$, as the field increases,  ${\partial
\tilde\mu}/{\partial B}$ sharply drops, then reaches a minimum,
and  saturates, or even increases at larger fields. The value of
$\partial\tilde\mu/\partial B= -\partial m/\partial n$ at the
minimum is approximately $-0.7\,\mu_B$, indicating almost full
spin polarization; at even lower densities it reaches a value of
almost $-\mu_B$. In the metallic regime, ${\partial
\tilde\mu}/{\partial B}$ changes sign. In Fig.\,\ref{Fig:dmdB} we
plotted ${\partial \tilde\mu}/{\partial B}$ for  $n=2\cdot
10^{11}{\rm cm^{-2}}$, the density at which it is maximal.
Although the saturation value at this density is much smaller,
about $0.15 \mu_B$, the saturation field $B^*$ is similar to the
one for $n=5\cdot 10^{10}{\,\rm cm^{-2}}$. Note that $B^*$ is
small, $\approx 0.6$\,T, well below the polarization field    for
a non-interacting system  ($\approx 11$\,T), and  for the
interacting one ($\approx 2.7$\,T), both  estimated  for $n=1\cdot
10^{11}{\rm cm^{-2}}$ \cite{PGK2001,pudalovsdh}. Such small $B^*$
implies an existence of a relevant energy scale, much smaller than
the bare, and  even interaction renormalized Fermi energy.

The decrease of $|{\partial \tilde\mu}/{\partial B}|$ at the
lowest temperature above $B^*$  can be attributed to the finite
thickness contributions: the slope of the solid line in
Fig.\,\ref{Fig:dmdB} reasonably agrees with the predicted FT
contributions for this density. As temperature grows, the
nonlinearity smears out, and its onset shifts to higher magnetic
fields\,\cite{Rcndmt}.

To explain qualitatively the results, one needs to invoke, besides itinerant electrons, states, localized either within the potential well\,\cite{DG}, or in its close vicinity. These states can be recharged at the frequency of the measurements,  and contribute to $\partial \mu_{2D}/{\partial B}$, but would not contribute to Hall conductivity or to Shubnikov-de Haas oscillations. The fact that the zero density gate voltage $V_0$  (i.e., the so called ``threshold voltage''), determined from extrapolation of the Hall or Shubnikov-de~Haas measurements, increases with reduction of electron mobility\,\cite{AFS} indicates the existence of such states.

On the other hand, the simplistic two-liquid model,  in which the non-interacting localized states exist
independently of the interacting 2DEG, suggested in\,\cite{DG}, cannot explain such a low value of $B^*$, since the characteristic field, $k_B T/\mu_B$, for polarization of a single spin (or non-interacting non-degenerate gas)  at 1.7K is 4.2T, much bigger than $B^*$. We therefore believe, that any explanation of the presented results must involve interactions. Note, that the underling mechanism should be different from the interactions between localized spins, discussed in\,\cite{BhatReview}, since such interactions lead to antiferromagnetic coupling, and slower than $1/T$ divergency of the susceptibility, in contrast with our observations \cite{Rcndmt}. We infer that coupling mediated by itinerant electrons should play a role.

Besides localized states, one can think about interaction-enhanced FT contributions. Indeed,  the estimations above ignored interactions. The interaction energy, however, is comparable to the energy difference between the levels of spatial quantization in the well; e.g., for $n=10^{11}{\rm \,cm^{-2}}$ both are about 10~meV. It is, therefore, not \emph{a priori} clear that the interactions can be ignored in calculations of $\bar z_0(B)$. One can envision a scenario, in which spin polarization leads to a change in $\bar z_0$. Note that $\delta \bar z_0\approx 1{\rm \AA}$ would suffice to contribute one Bohr magneton to $\partial \tilde \mu/\partial B$. It would be, however, quite a coincidence if such a mechanism conspires to contribute  $\partial \tilde\mu/\partial B\approx -\mu_B$ at the saturation.

In conclusion, we succeeded in measuring  the magnetic field
dependence of $\partial \mu_{2D}/\partial B$ over a density range
much wider than was accessible earlier, from metallic to deep in
the insulating region.  The dependence of $\partial m/\partial n$
on $B$ is found to be strongly nonlinear at low temperatures,
already in magnetic fields much smaller than the characteristic
fields expected for the 2D electron gas. These  results point to a
strong coupling between the itinerant and localized states in high
mobility Si-MOSFETs.

We would like to thank A.\,Efros and A.\,Finkel'stein for discussions. M.R. is thankful to the Israel Science Foundation, Binational Science foundation and Russell Berrie Nanothechnology Institute for financial support. A.Yu.K. and V.M.P acknowledge support by Russian Academy of sciences, RFBR, and Russian ministry for education and science (under contracts No
02.552.11.7093,  14.740.11.0061, P2306, P798,  P1234).


\begin{thebibliography}{qq}
\bibitem{AFS}  T.~Ando, A.~B.~Fowler, F.~Stern, Rev. Mod. Phys. {\bf 54}, 437 (1982).
\bibitem{Stoner}E.~C. Stoner,
Rep. Prog. Phys. {\bf 11} 43 (1947).
\bibitem{BhatReview}R.~N.~Bhatt, Physica Scripta {\bf T14} 7-16, (1986).
\bibitem{simonyan}
D.\,Simonian, S.\,V.\,Kravchenko, M.\,P.\,Sarachik, and V.\,M.\,Pudalov, Phys. Rev. Lett. {\bf 79},
2304 (1997); V.\,M.\,Pudalov, G.\,Brunthaler, A.\,Prinz, and G.\,Bauer, JETP Lett. {\bf 65}, 932  (1997).

\bibitem{SKDK} A.\,A.\,Shashkin, S.\,V.\,Kravchenko, V.\,T.\,Dolgopolov, T.\,M.\,Klapwijk,
Phys. Rev. Lett. {\bf 87}, 086801 (2001).
\bibitem{VZ2} S.\,A.\,Vitkalov, H.\,Zheng, K.\,M.\,Mertes, M.\,P.\,Sarachik, T.\,M.\,Klapwijk,
Phys. Rev. Lett. {\bf 87}, 086401 (2001).

\bibitem{pudalovsdh} V.\,M.\,Pudalov, M.\,E.\,Gershenson, H.\,Kojima, N.\,Butch, E.\,M.\,Dizhur,
    G.\,Brunthaler, A.\,Prinz, and G.\,Bauer, Phys. Rev. Lett. {\bf 88}, 196404 (2002).
\bibitem{PGK2001} V.\,M.\,Pudalov, M.\,E.\,Gershenson, and H.\,Kojima, arXiv:cond-mat/0110160; arXiv:cond-mat/0401396v2.
\bibitem{Prus2003} O.\,Prus, Y.\,Yaish, M.\,Reznikov, U.\,Sivan, and V.M.\,Pudalov, Phys. Rev. B {\bf 67}, 205407 (2003).
\bibitem{shashkintd} A.\,A.\,Shashkin, S.\,Anissimova, M.\,R.\,Sakr, S.\,V.\,Kravchenko, V.\,T.\,Dolgopolov,
    and T.\,M.\,Klapwijk, Phys. Rev. Lett. {\bf 96}, 036403 (2006).
\bibitem{commentonC} The exact equation for the measured capacitance is ${\tilde c}=c_0(1+\frac{\partial c_0}{\partial n}\frac{n}{c_0})/\left(1-\frac{\partial c_0}{\partial n}\frac{n}{c_0}+\frac{c_0}{e^2}\frac{\partial \mu_{2D}}{\partial
n}\right)$.
\bibitem{Rcndmt} N. Teneh, A. Yu. Kuntsevich, V. M. Pudalov, T. M. Klapwijk, and M. Reznikov, arXiv:0910.5724v1 (2009).
\bibitem{comment_R_S} M.\,Reznikov, U.\,Sivan,  arXiv:cond-mat/0410409
\bibitem{Dolgopolov_skin} S.\,I.\, Dorozhkin, A.\,A.\,Shashkin, N.\,B.\,Zhitenev and B.\,T.\,Dolgopolov, Pis'ma Zh. Exp. Teor. Fiz. {\bf 4}, 189 (1986); JETP Lett. {\bf 44}, 241 (1986)
\bibitem{DG} A.\,Gold, V.\,T\,Dolgopolov, J. Phys.: Condensed Matter {\bf 14}, 7091 (2002).


\end{thebibliography}
\end{document}